# Winds Through Time: Interactive Data Visualization and Physicalization for Paleoclimate Communication

David Hunter
ATLAS Institute
University of Colorado at Boulder
david.hunter@colorado.edu

Pablo Botin
University of Colorado at Boulder
pablo.botin@colorado.edu

Emily Snode-Brenneman
UCAR and NSF NCAR
emilysb@ucar.edu

Amy Stevermer
UCAR and NSF NCAR
asteverm@ucar.edu

Becca Hatheway
UCAR and NSF NCAR
hatheway@ucar.edu

Dillon Amaya
NOAA
dillon.amaya@noaa.gov

Eddie Goldstein
Science Communicator
eddie@eddiegoldstein.com

Wayne A Seltzer
ATLAS Institute
University of Colorado at Boulder
wayne.seltzer@colorado.edu

Mark D Gross
ATLAS Institute
University of Colorado at Boulder
mdgross@colorado.edu

Kris Karnauskas
University of Colorado at Boulder
kristopher.karnauskas@colorado.edu

Daniel Leithinger
Cornell University
daniel.leithinger@cornell.edu

Ellen Yi-Luen Do
ATLAS Institute
University of Colorado at Boulder
ellen.do@colorado.edu

## Abstract
We describe a multidisciplinary collaboration to iteratively design an interactive exhibit for a public science center on paleoclimate, the study of past climates. We created a data physicalisation of mountains and ice sheets that can be tangibly manipulated by visitors to interact with a wind simulation visualisation that demonstrates how the climate of North America differed dramatically between now and the peak of the last ice age. We detail the system for interaction and visualisation plus design choices to appeal to an audience that ranges from children to scientists and responds to site requirements.

## Introduction

We share our collaborative design process to create a hands-on interactive exhibit for visitors to learn about paleoclimate, the study of past climates on earth, in particular the effect ice sheets had on climate during the last ice age. The exhibit was commissioned by the UCAR and NSF NCAR for their public science centre, the Mesa Lab, in the Rocky Mountains.

Our exhibit synthesizes tangible and visual forms of data representation through projecting a reactive simulation visualization on to a tangible objects that function as a physicalisation and an interface. In this approach we analyzed what data could be "static" and represented through physicalization and what could be "dynamic" and needed to be an animated/generative visualization. We used a depth camera to bridge real world physical visitor interactions with a wind simulation and visualization.

The complexities of atmospheric science necessitated an innovative wind simulation that was lightweight enough to run on consumer computers in real-time but convey the subtlety of wind movements at the continental-scale. Through iteration we refined a visualization strategy for large quantities of dynamic simulated data. Our exhibit had to be robust to meet a number of on-site requirements, as well as the challenges of delivering a visitor experience that balanced the need for nuance and accuracy in scientific messaging with engagement for a broad audience ranging from young children to adult scientific experts.

The project arose through a multi-disciplinary network of scientists, educators, students, and science center staff, who in a sustained and collective way transmuted scientific research into designed artefact, passing new knowledge from research office to classroom to public exhibit, and into the hands of visitors.

collective care

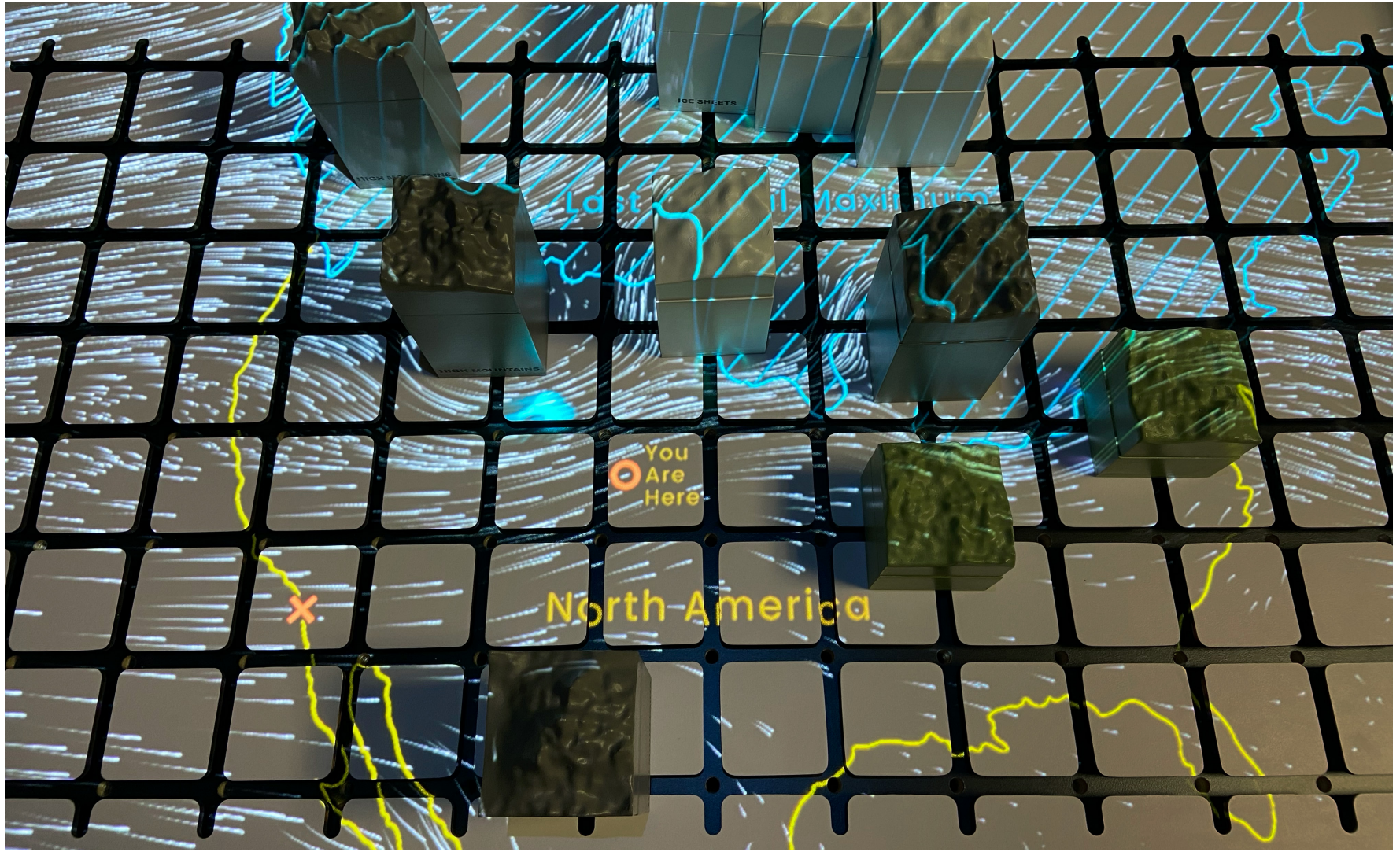

### Winds Through Time
In this exhibit, visitors can move blocks (representing mountains and ice sheets) across a tabletop, altering wind pattern projected onto the tabletop and learn about paleoclimate through tangible interaction with simulation data.

### Wind Visualization
Wind movements are visualized as particles and storms, represented by an icon, that flow across the tabletop and around the blocks. We use a custom-developed Computational Fluid Dynamics simulation to replicate large-scale wind patterns.

### Data Physicalization
The height and color of each block is a physicalisation of categorical data: ice sheets, high mountains, and low mountains. Block height is sensed by a depth camera and input to the wind simulation to affect movements.

### Map Context
To situate and inform visitors we designed a minimalistic map provides a minimal visual with recognizable outlines, labels and icons to identify regions and features, without obstructing the wind simulation.

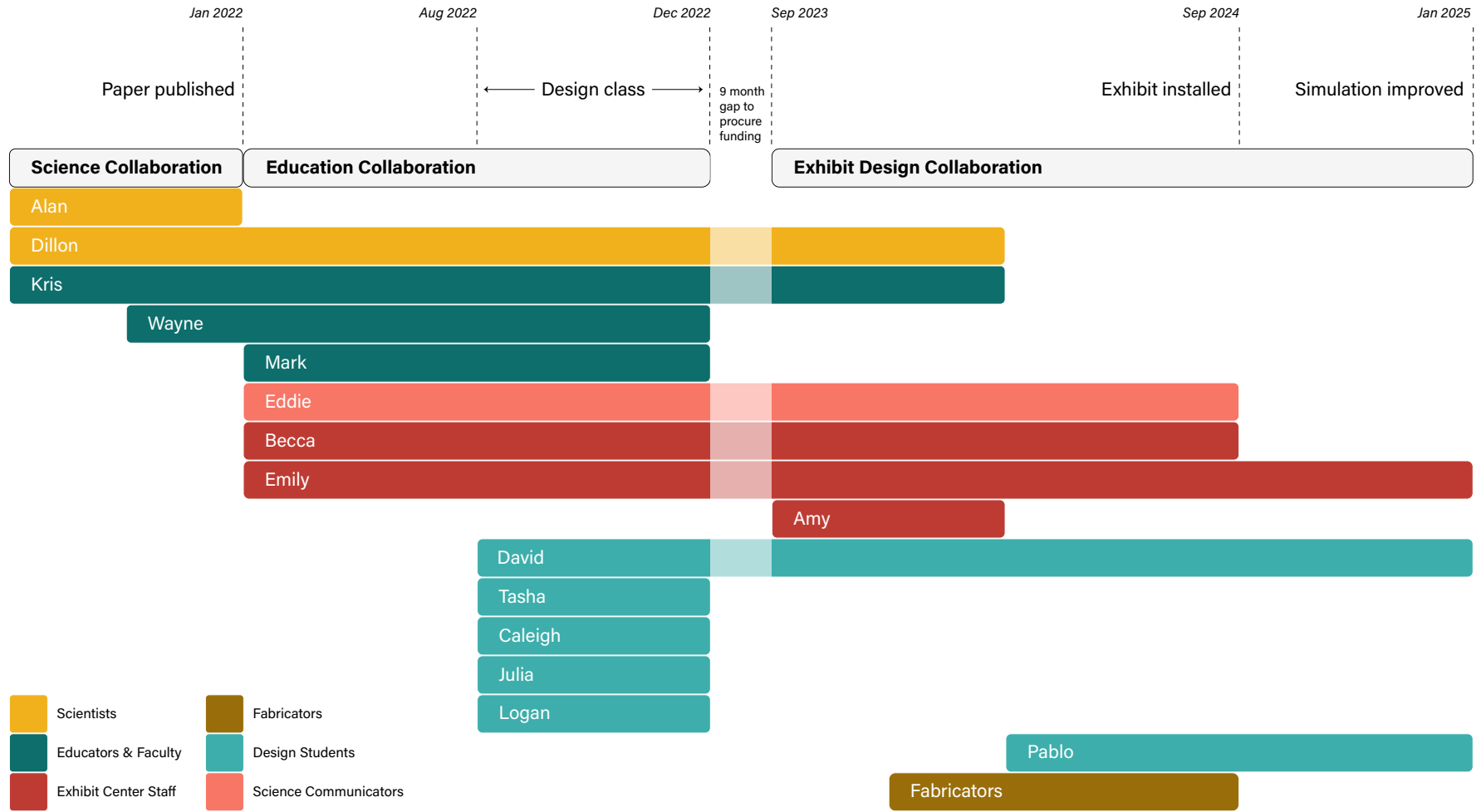

## The Collaborative Approach

This project began with a scientific paper [2] investigating how during the Last Glacial Maximum the large height of ice sheets and their albedo (reflectivity), shifted winds and storms coming from the Pacific over North America to the south south. This made the South West 20,000 years ago a much wetter climate than we experience today.

Through personal connections this research was incorporated into a class on designing a science exhibit; an opportunity for scientists, educators, and design students to work together creating prototype exhibits for the Mesa Lab. The project in this pictorial was selected to be turned into a permanent exhibit, and once funding was secured a team was assembled comprising of David, a graduate student designer, Eddie, a science educator, Dillon, a climate scientist, and from Mesa Lab Becca, a curator, Emily, an exhibit manager, and Amy, an exhibit designer. The multidisciplinary team worked beyond their domain expertise, developing ideas, iteratively refining designs, and collaboratively making decisions ,all while listening to each other's expertise.

## The Initial Prototype: White Mountain, Green Mountain

Our group of five students decided to create a tangible user interface for visitors to explore how relief (mountains and ice sheets) and albedo (reflectivity) changed wind and climate during the last ice age. We based our design on the AR Sandbox [4] where users can modify the contents of a sandpit, a Kinect depth camera captures a depth image of the sandpit and graphics are projected over the sandpit. We followed the AR Sandbox construction guide, but swapped sand for blocks to reduce mess. We tested two block concepts, either with many smaller blocks that users could sculpt into landscapes, or much larger blocks representing ice sheets and mountains. Green blocks indicated mountains and white blocks indicated ice sheets.

We created a custom wind simulation using Processing that moved particles according to repulsive forces (explained in more detail on page 7). Initially we tried flow fields to determine wind direction but the wind did not flow correctly when directed past 180°. We modelled the albedo of ice sheets by detecting the color of the blocks, where white areas shifted the wind vector further south. Particles accumulated atmospheric moisture and when reaching "saturation" point either by passing over high blocks or after accumulating above a threshold the particle would release rainfall. The ground was divided into a grid which visualized how much rain had fallen on that area. We experimented with different visualisations of the wind and rain, and tested our simulation with subject matter experts from the Mesa Lab, science education. Our final prototype was displayed alongside other class group prototypes. The Mesa Lab team selected our prototype to be improved and become part of their permanent display.

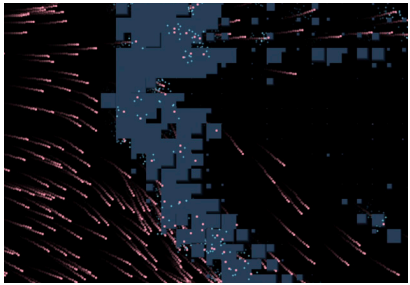
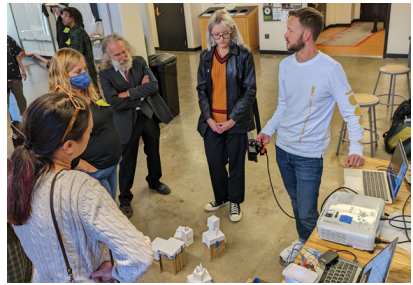
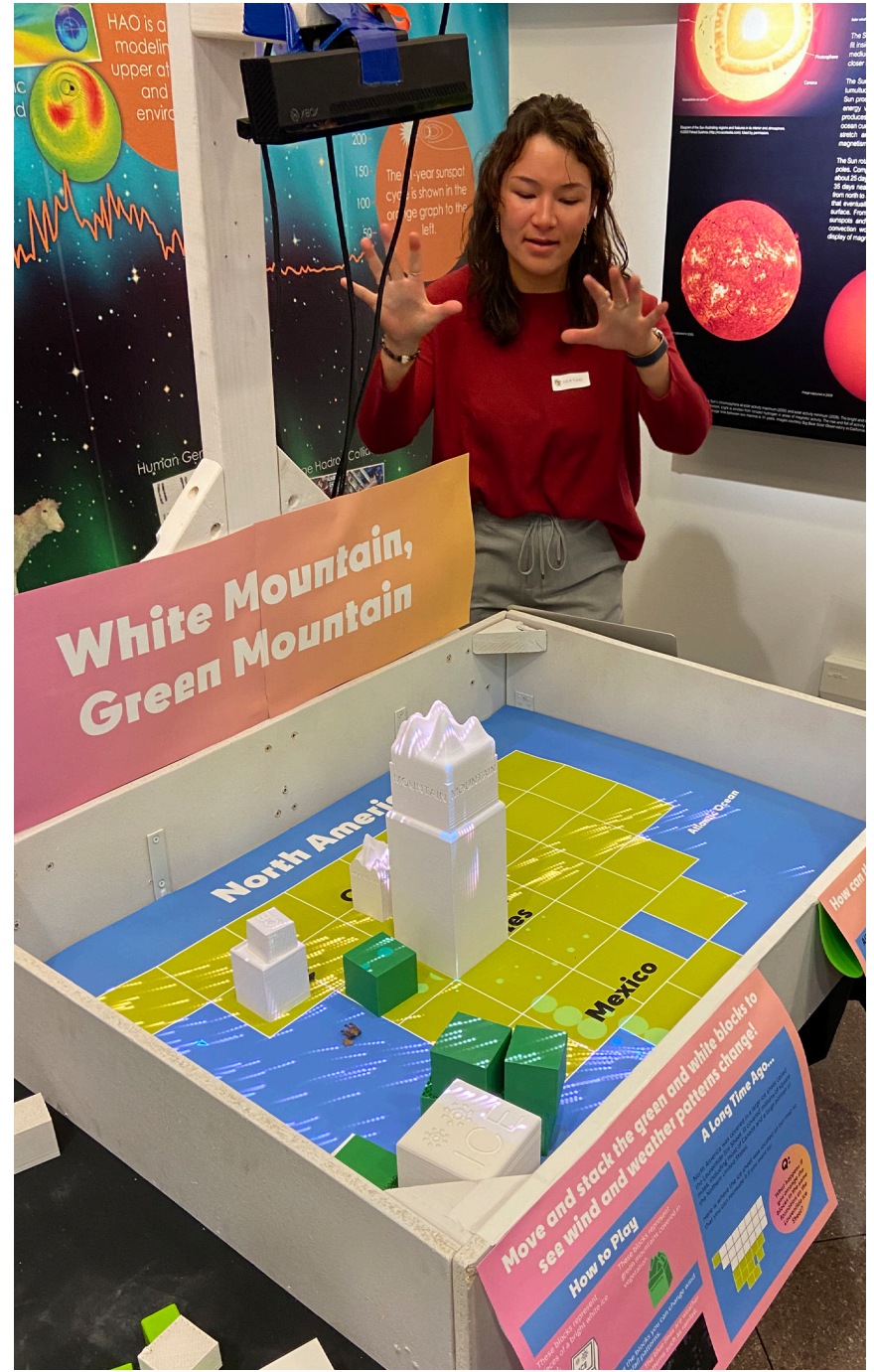

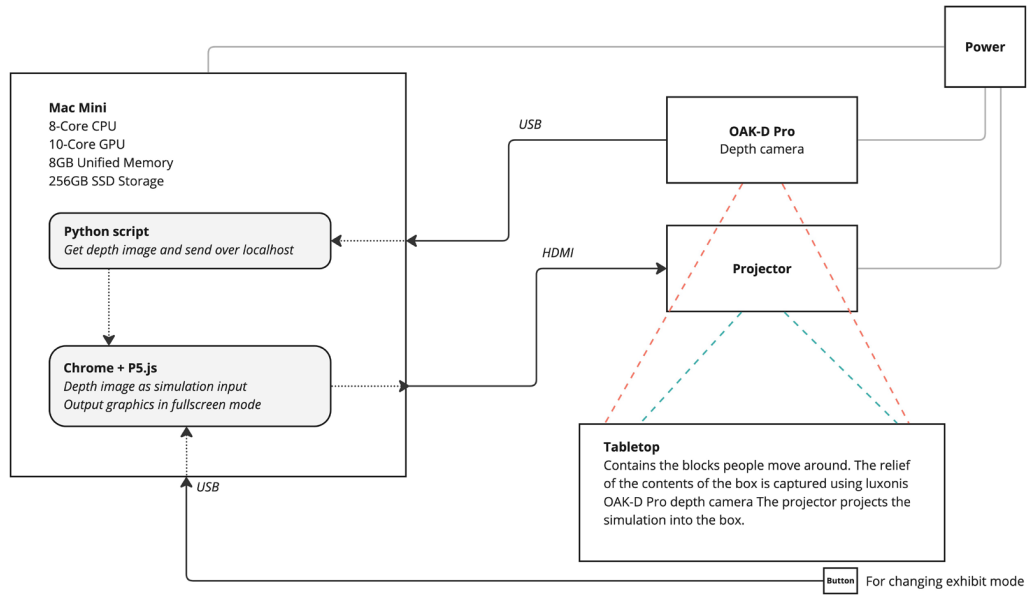
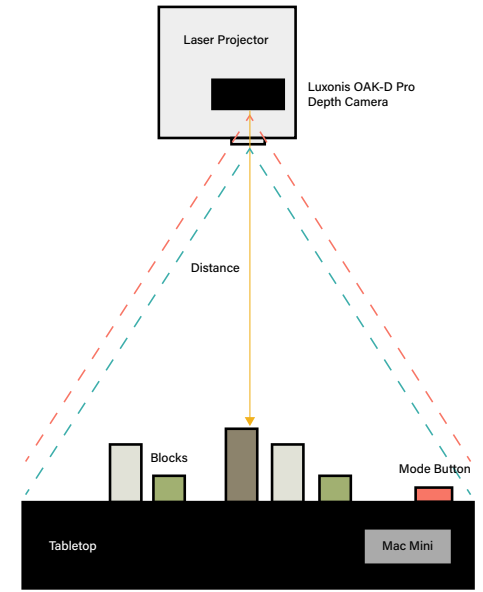

### The Winds Through Time System

For the permanent exhibit we designed a new system to last 5-10 years using easily replaced and well supported technology. We opted for a Luxonis depth camera, and rewrote the simulation in P5.js to avoid managing Java. The team decided to cut the albedo and rain elements from the simulation and focus purely on wind movements—with a broad audience, there was a risk that too many simulation variables could lead to confusion on the science. The team wanted visitors to understand one factor well, rather than misunderstand several.

With the new system, visitors move blocks representing relief (ice sheets, mountains) around the tabletop to construct mountain ranges and large ice sheets. The sensing system depth camera looks down on the tabletop to capture a 2.5D representation of the tabletop and blocks as a depth image. Areas of high relief (closer to the camera) are passed to the simulation as areas to "avoid", driving the particles to react and move around those areas. The visualization is projected from a ceiling mount. We continued with the repulsive forces concept (detailed on p.7), and later upgraded to a computational fluid dynamics simulation (detailed on p.13). Particles generate from the left edge, to match the prevailing westerly winds, but some particles start at random places on the tabletop for a more even dispersion. The depth camera is calibrated to only read depths within a certain distance range with millimeter accuracy, to ignore hands or heads hovering over the table. This improves user experience by preventing unwanted simulation effects from visitor interactions.

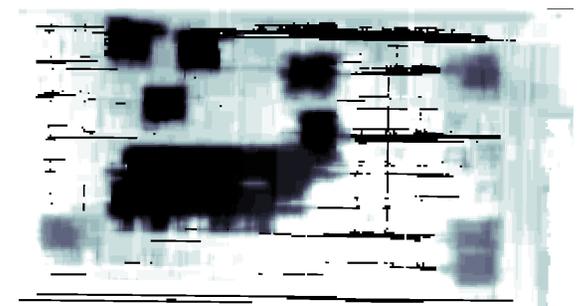

### Depth Image

Above shows an example depth image of the tabletop sent to the simulation. The very dark clusters show tall blocks while the lighter areas are shorter blocks. By setting thresholds with color values the simulation can react at the pixel level to object depth and therefore complex arrangements of blocks by visitors, including position and rotation. Noise can be filtered out by calibrating the depth camera in the python script.

## Simulating and Visualising Wind Movement with a Partical System using Repulsive Forces

We developed a concept of using repulsive forces to represent the interactions between wind and large relief features such as mountains and ice sheets, when viewed from a top-down view of a landscape.

Large relief features force particles representing wind to avoid going over the large features and instead moving in a less resistant direction around the feature. With multiple features the particles can effectively be "steered" around or between features, simulating how wind might travel in a valley or around a mountain. This provided a reasonable model of large scale wind movements but lacked nuances like vortices or eddies that form in the real world natural and complex systems.

To create the particle trails effect, an opaque (instead of solid) black rectangle at 10% opacity is drawn as a background leaving echoes of the previous ten renderings partially visible.

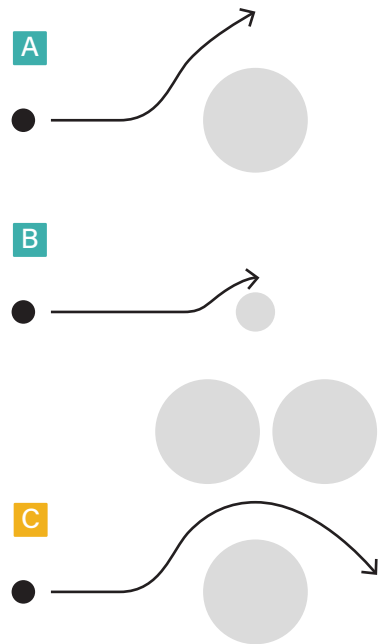

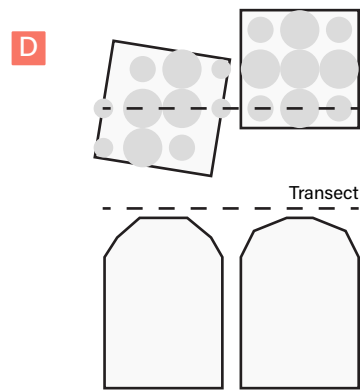

**A, B** The size of the repulsive forces changes a particle's deflection.
**C** Forces can steer a particle.
**D** Arrangements of shapes can be described by grids of forces. The size of force equates to the object height.

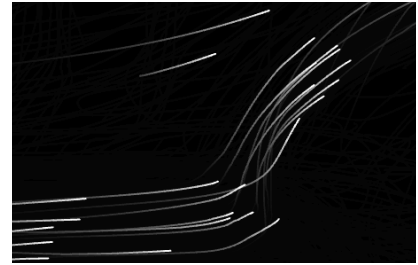

Our selected style used small white particles with long trails, familiar to wind visualizations on weather maps, and this also contrasted well when projected even with strong daylight.

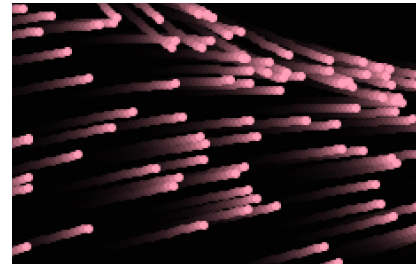

For our initial prototype we used large colored particles with short trails. However, the color proved confusing in relation to wind to some viewers. Larger ellipses slowed the rendering.

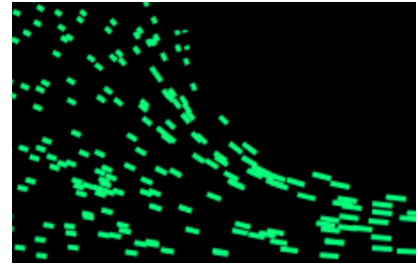

We also explored particles without trails, instead the length of the line was determined by the velocity of the particle. Ultimately, the team preferred the fading effect of trails.

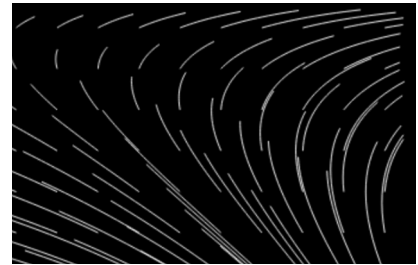

A vector field of lines in fixed positions that rotate and extend dependent on wind direction and magnitude. While an effective representation of data, it did not feel familiar to the team.

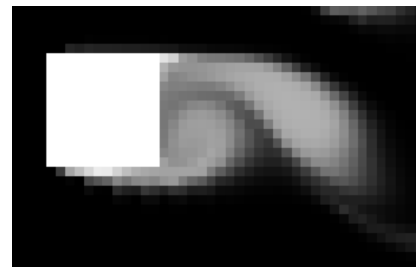

After improving the simulation to use Computational Fluid Dynamics, a "smoke" style became an option, but it was easy to produce too much smoke and overwhelm the visualisation.

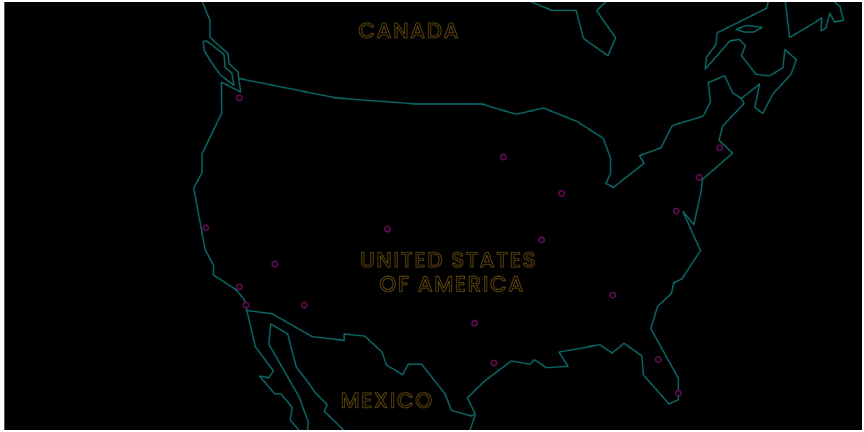
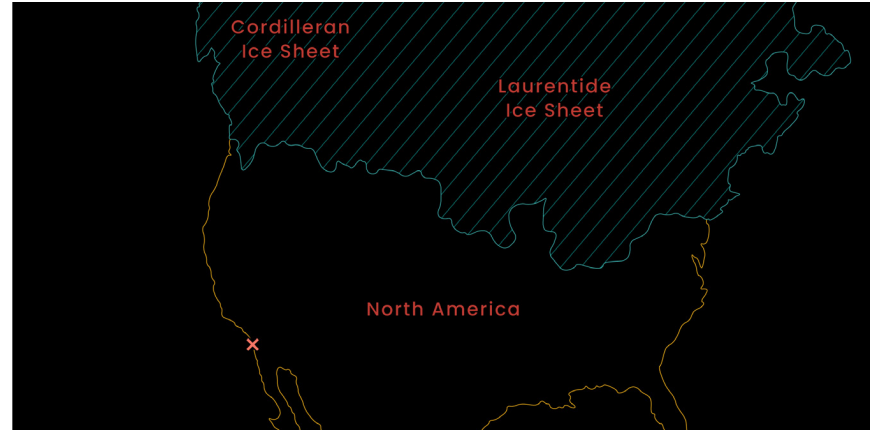
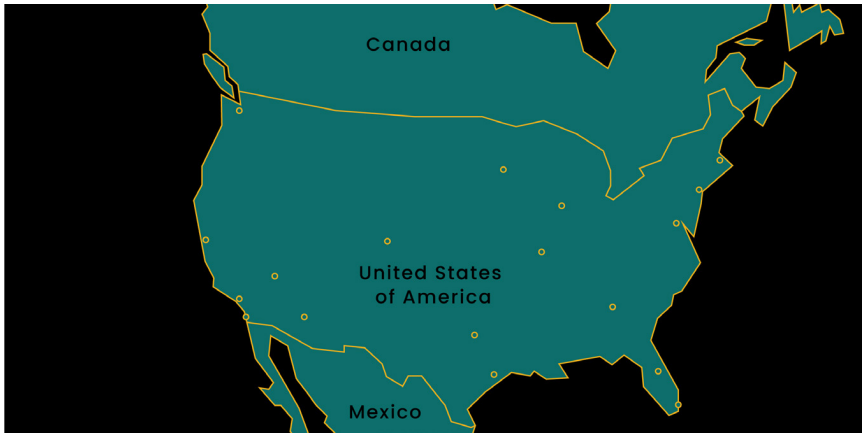
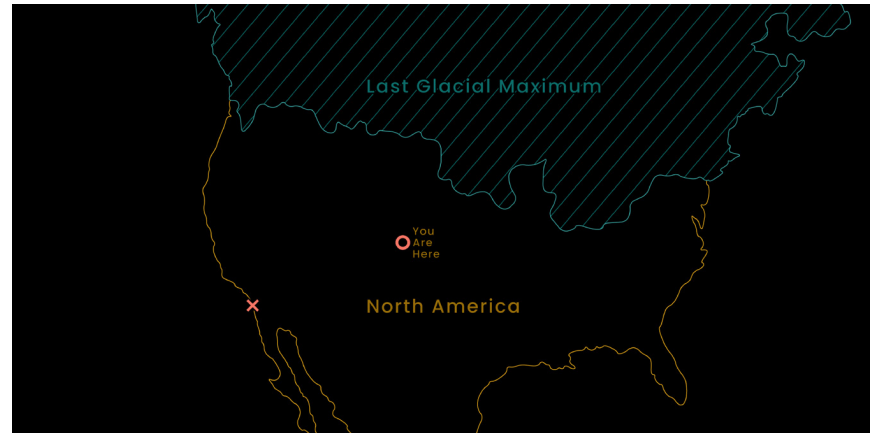
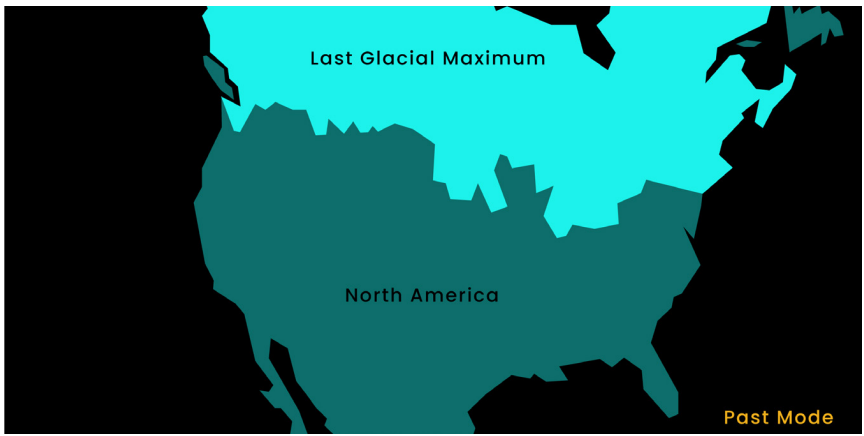

## Map Design

Early on we decided to fix the bounds of the map to the North American continent, in line with the scientific paper. A sample of our map design iterations are shown, exploring color, filled or outlined and angular or rounded tracing of features. We experimented with what labels to include and how to apply colors semantically: should we color according to type of feature or distinguish labels as a separate color? Initially we used country labels and borders, but opted to color and label according to key geographic features that relate to the story of paleoclimate, a time when cities and country borders did not exist. We added a marker to help situate visitors and an X target to steer storms towards when mountains and ice sheets are in the correct position.

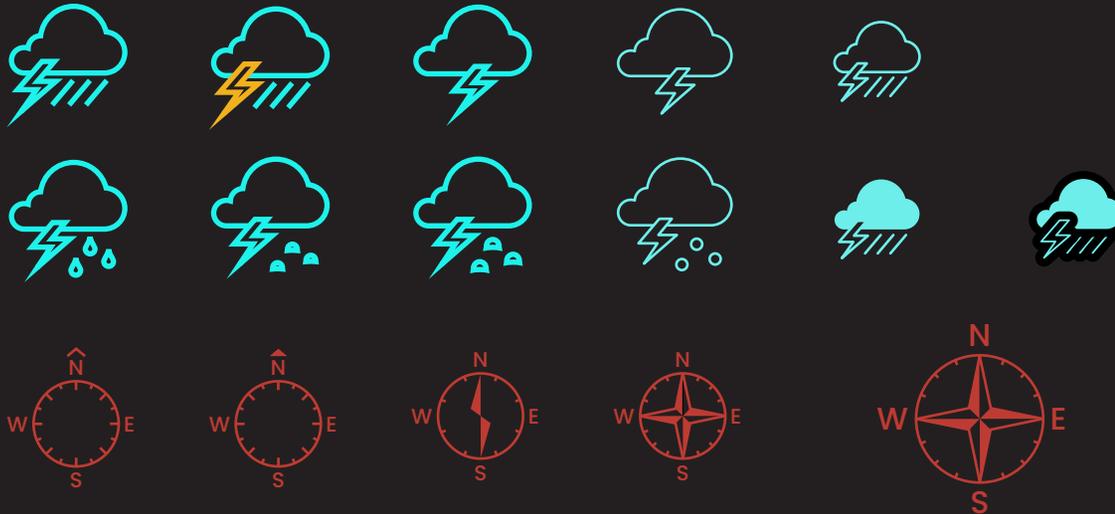

| Land Outline | Location Icon | Glacial Label / Ice Sheet Hatching | Ice Sheet Extent Outline |
|---|---|---|---|
| Land Label | Mode Label / Compass Icon | Ice Sheet Label | Storm Icon |

## Color Scheme
Bright and distinct colours were chosen for the final map and UI graphics to standout in a room with natural light diluting the projection strength. Tundra colors pointed to ice features and those were contrasted with gold for land features. A light and dark tonal variation of color indicate a thematic connection.

## Storm Icon
Icons were designed iteratively for readability when projected. The storm icon would be moving across the tabletop with the wind particles, and therefore had to be readable when moving.

## Compass Icon
The compass, as is tradition on a map, was to orient the user; important considering the impact of hemispheres have on the behaviour of wind at a global level. We increased the size of the compass and added a distance marker to help visitors understand the scale of features and movements. We used both metric and imperial units for easy comprehension.

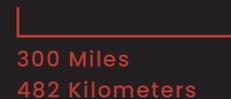

300 Miles
482 Kilometers

## Location Icon
The X target implies where you are aiming the wind to go towards, while the O marker indicates the location of the Mesa Lab in North America.

**Moving Mountains Mode**

**Ice Age Mode**

## Typography
We selected Poppins for the typography on the tabletop interface as an elegant and legible choice for projection.

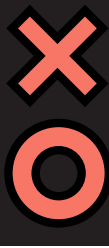

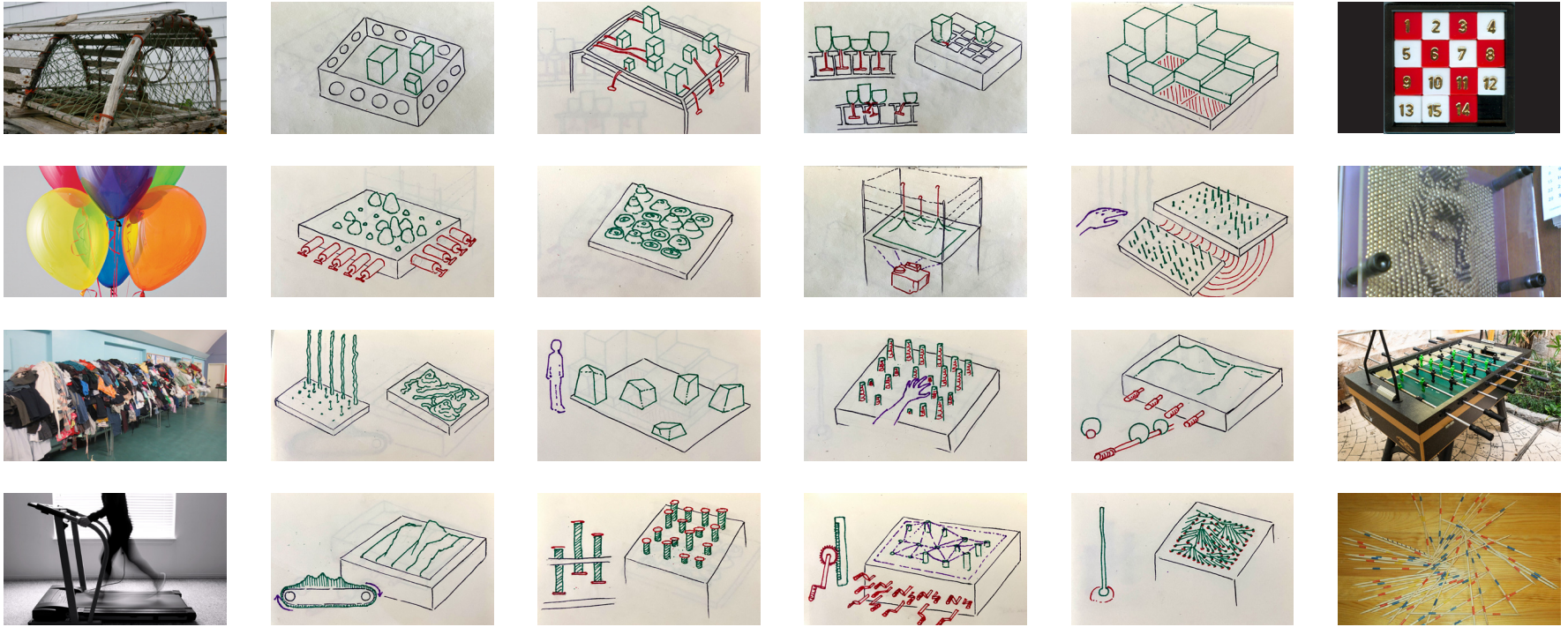

## Designing a Tangible User Interface (TUI)

For the permanent exhibit to be viable for public display, a set of specifications had to be met:
- Elements of the exhibit must not leave the exhibit area
- The exhibit must not create any mess
- The exhibit should be low-maintenance
- The exhibit must be durable for 5-10 years

The reasons and justification for these specifications were:
- Low staffing: staff cannot tend to the exhibit all day or clean up mess
- Health & safety: mess is a trip hazard for visitors and staff
- Active workspace: exhibits cannot disrupt workers in the building

How could the TUI be "bounded" so it could not be removed from the exhibit area? Inspired by every day real world examples of tangible interfaces and interactive mechanisms, we sketched 16 ideas for "bounded" TUIs shown above.

We looked at existing research on TUI [3] and exhibit design [8], TUI design and frameworks [7, 9], to inform our conceptual approach on tokens, constraints, interactive surfaces, and shape-changing displays [1]. Much of this research is underpinned by the study of affordances [5, 6, 10], the physical and cultural aspects of interaction with an object or design.

Based on the initial idea generation and sketching, and existing TUI literature, we developed a set of parameters to define a successful TUI for the exhibit:
- Method of bounding
- Freedom of movement
- Resolution of Tokens
- Stackability of Tokens
- Freedom of Manipulation
- Method of Interaction
- Discreteness of Token

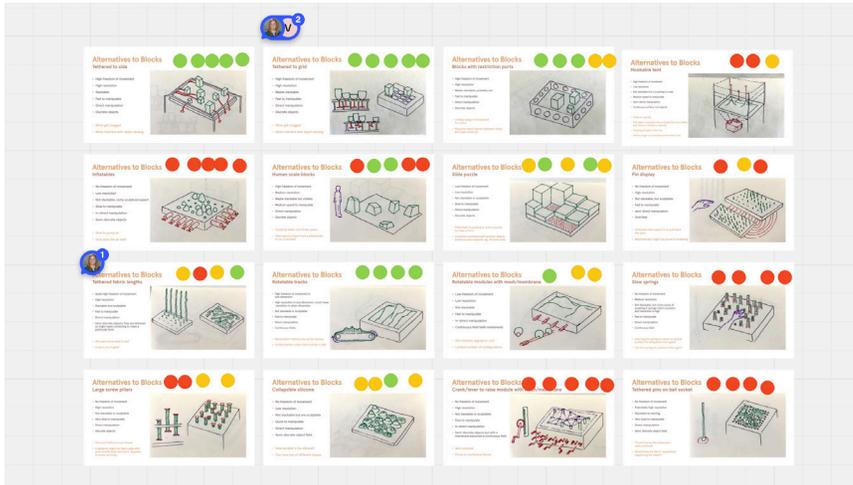

### Evaluating our Ideas through Dot Voting

After ideas were sketched we held a team workshop to help whittle down the ideas for a "bounded" tangible user interface. Drawings were compiled with descriptive notes on how they met the parameters which would lead to success. We held a discussion of each idea to clarify any questions, and then conducted a dot voting exercise, where each member could vote whether it was a good idea (green), needed more discussion (yellow), or not a good idea (red). This voting gave us a good map for which ideas to proceed with prototyping. The successful ideas leaned towards blocks that could be moved as freely as possible, rather than fixed grids of objects that the height could be adjusted or sculptable surfaces/objects.

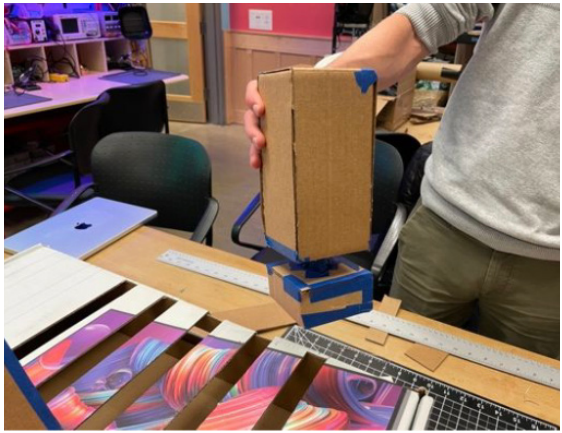
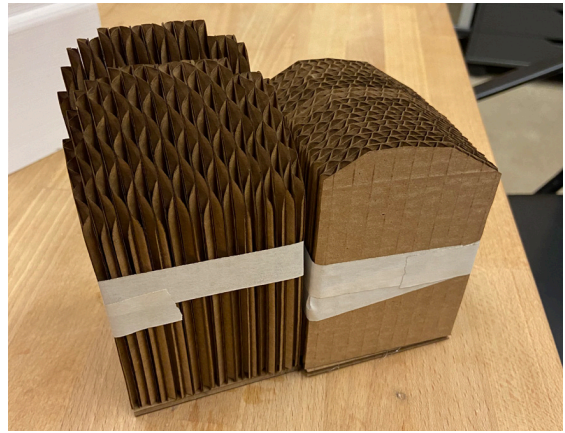
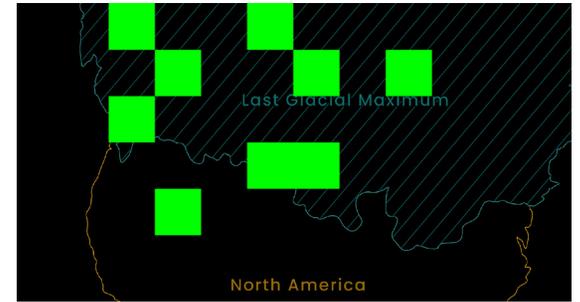
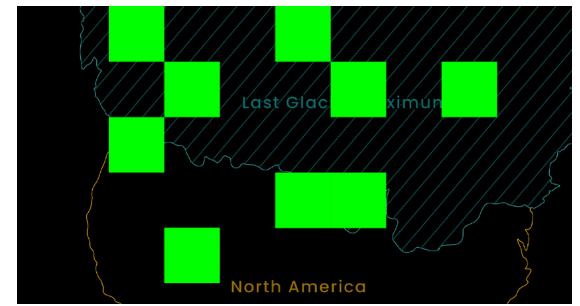

### Block and Mechanism Prototyping

Using scrap cardboard we prototyped a basic versions of the mechanism to "bound" the blocks to take to the fabrication company. We also prototyped the fabric concept, but this proved too unfamiliar.

For the blocks we prototyped cardboard blocks at different sizes to investigate how easy was it to grasp and move for adults and children? This required balancing ergonomics with how many blocks to have: We needed enough to fill the area of the last glacial maximum on the map, but not so many that it would be tedious to move 100 blocks, but not so few that the blocks were too big to grasp or didn't allow for a wide range of arrangements on the tabletop. We decided on between 10-20 blocks as being suitable. Prototyping allowed us to check various block heights could be differentiated by the depth sensor to produce the correct simulation effects.

### Checking Geographic Scale

We also checked if the proposed block size and grid resolution would match well to geographic scale against our map.

## On-Site User Testing

We spent an afternoon at the Mesa Lab with a low-fidelity prototype available for the public to test, noting observations and user feedback. This session included visiting school groups, families, and scientists who work in the building. The prototype consisted of a version of the simulation projected onto a cardboard panel on the floor at the size of the tabletop. A number of cardboard blocks were made at the intended dimensions with basic surface modulation to estimate how different types of blocks would look.

The goal of the session was to find out if the block size was comfortable to use, if map designs were recognizable, if the wind visualizations were interpretable, and if the behaviour of the simulation satisfied expert-scientists and non-expert visitors. Kids from the school group took part in a guided session with a facilitator who set the kids a task to direct the wind to points on the map, and could free-play with the prototype. Passing visitors and scientists were invited to try the prototype.

The block size proved to be successful in terms of graspability and the different heights. The labels on the blocks were helpful, however we had not color-coded the blocks at that point. The map was easily recognisable to visitors and we observed that less sections filled with color was better for seeing the wind simulation. Without the map some children interpreted the wind particles as shooting stars, so anchoring the experience geographically was important. Scientists liked the simulation but wanted to see more nuances in the way the wind reacted to the blocks.

People were interested in the technology, understanding how it worked, but they were more captivated by the simulation and exploring the effect of the blocks.

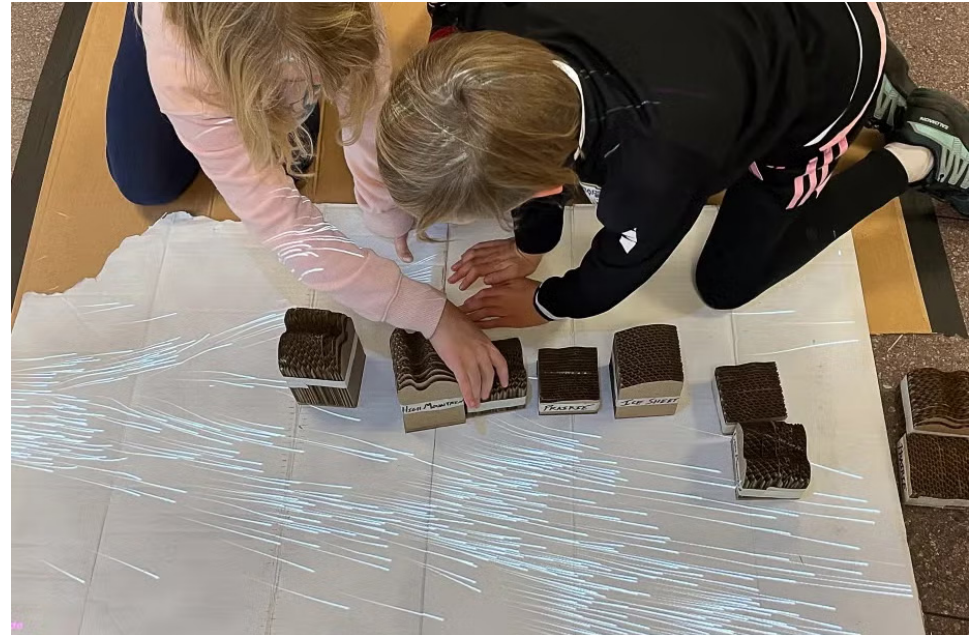
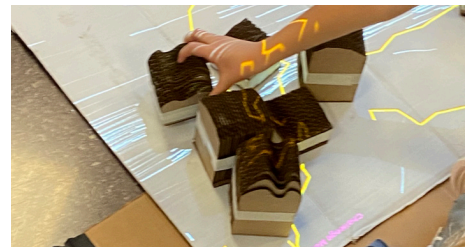
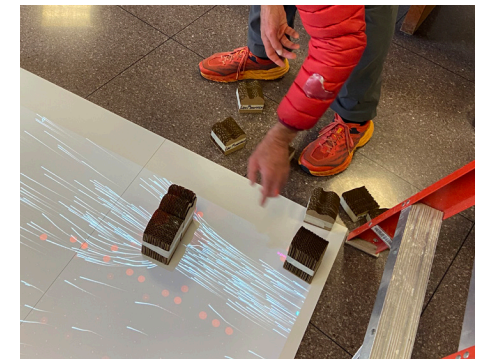
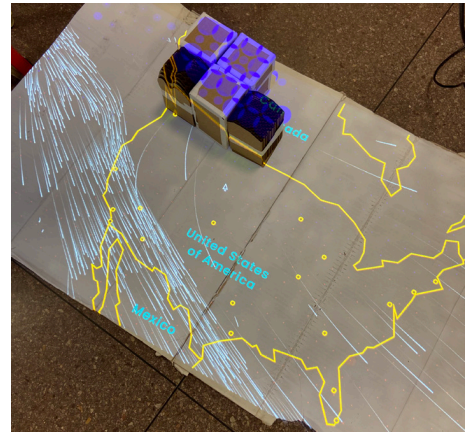
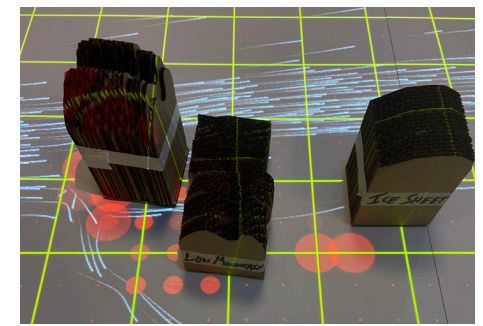

## Data Physicalisation: Designing the Blocks

Each block is designed to represent one of three types relief category: *low mountains* that have little impact on wind patterns at continental-scale, *high mountains* like 14'ers (14,000ft) in the Rocky Mountains that do influence wind patterns, and *ice sheets* like those found over North America during the last age that the science paper identified were a factor driving winds south in paleoclimate.

Blocks were modelled in Blender using Perlin noise to create a surface features: spiky high mountains, smoother low mountains, and undulating ice sheets. Experimentation was needed to get the right level of detail, feature height variation, and edge slope. The 3D printed surfaces were attached to wooden cuboids for the desired height, and painted a color differentiated by category. We took a categorical approach to aid comprehension for young visitors.

## Test 3D Prints

While iterating our landform blocks we 3D printed them to check how they looked at their true physical size rather than screen renders. Some designs looked better when rendered rather than printed.

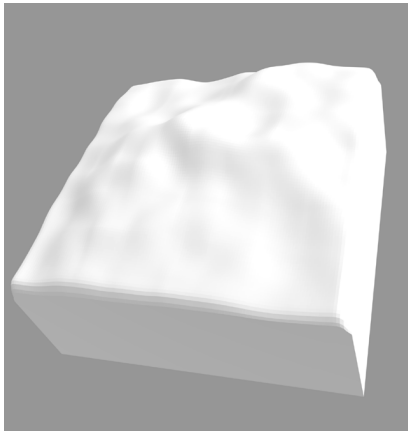
Ice Sheets

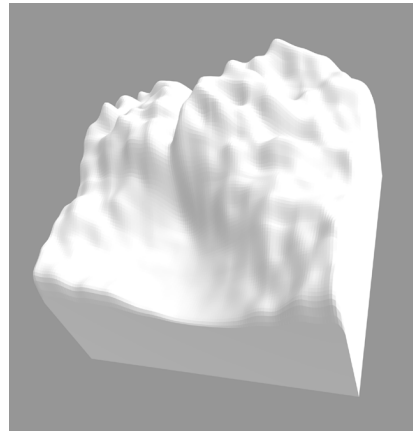
High Mountains

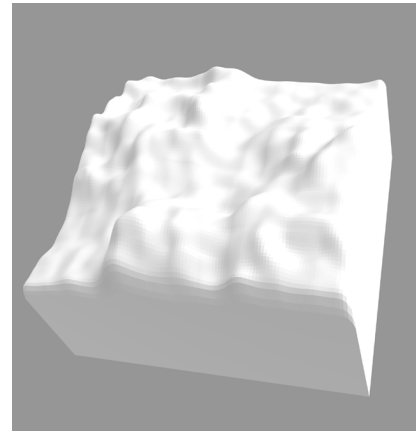
Low Mountains

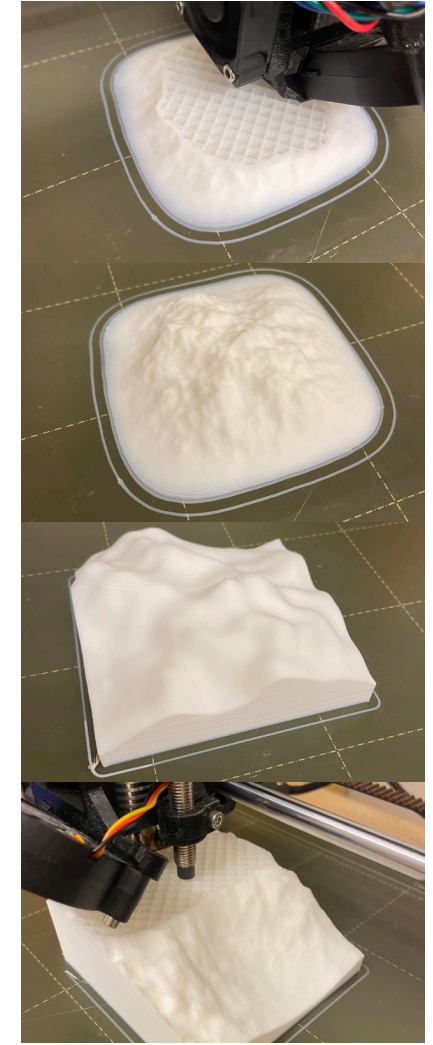

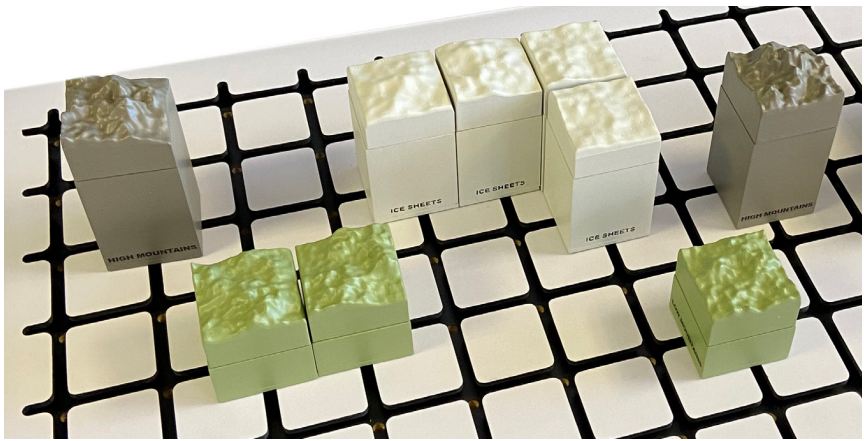

## Designing the Tabletop

The tabletop enables visitors to slide blocks from a storage area onto the projection area and control the number and type of blocks on the tabletop.

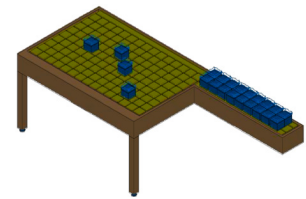

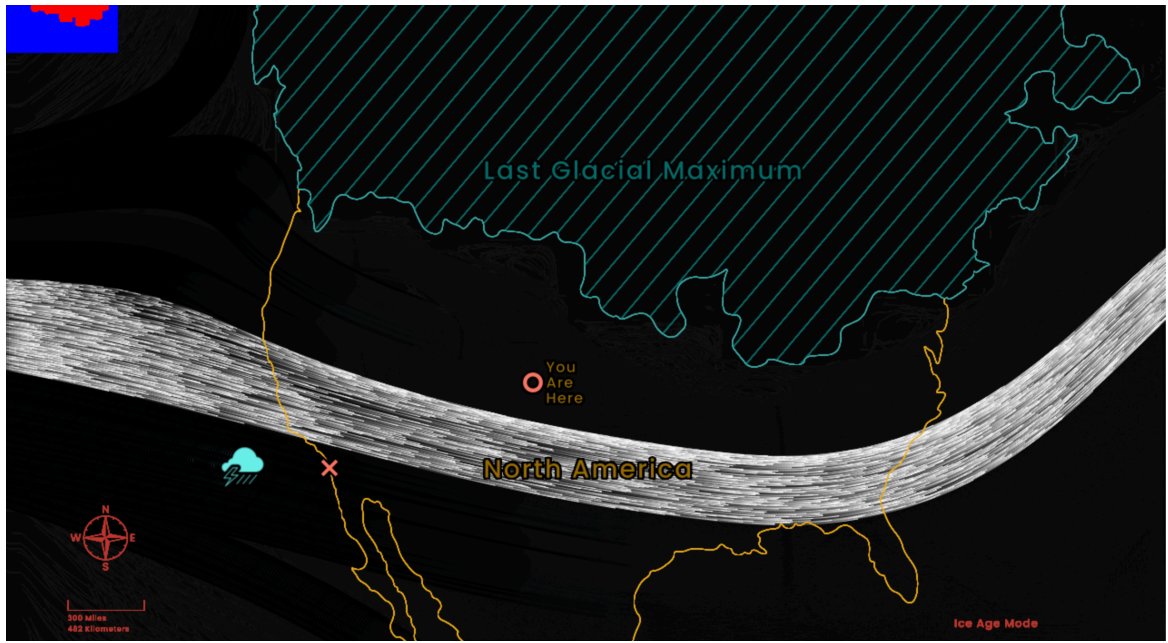

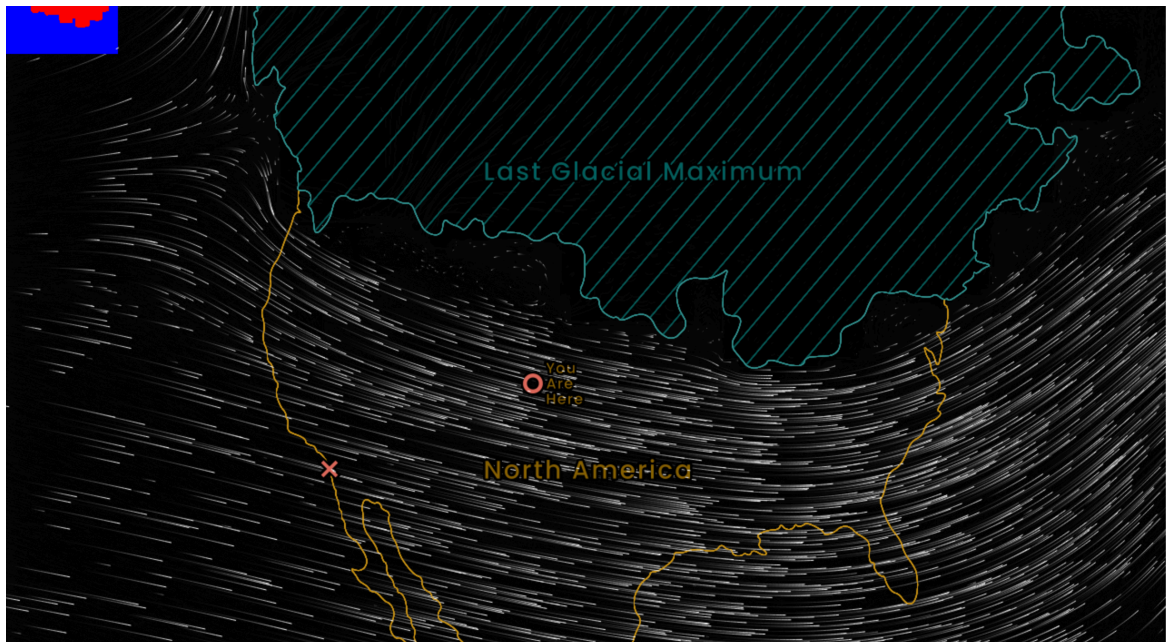

## Improving the Simulation

While the repulsive forces concept gave a reasonable model for wind moving across continental America, the wind behaviour lacked important nuances that the team and scientists would like to see in the simulation.

We recruited Pablo, a Computational Fluid Dynamics (CFD) expert, to work with David on a new simulation using CFD. However, CFD alone does not capture the full range of dynamics of wind movements at continental scale. The Coriolis effect, is one such important factor, where the Earth's rotation deflects air movements causing them to curve. The effect is more extreme at higher latitudes. The Coriolis effect causes winds in the northern hemisphere to bend to the right, and to the left in the southern hemisphere. Rossby Waves, large-scale undulating waves that can stretch for thousands of kilometres are formed in part due to the Coriolis effect (and conservation of potential vorticity). On the left you can see a Rossby Wave with a narrow jet of wind and below a wider less concentrated jet in a Rossby Wave.

Pablo customized the CFD equations to include the Coriolis effect. After a lot of testing we found a balance between the strength of the Coriolis effect and other factors like velocity and friction. The updated simulation produced a more convincing visual result with Rossby Waves and smaller nuances like vortices from wind moving around the blocks.

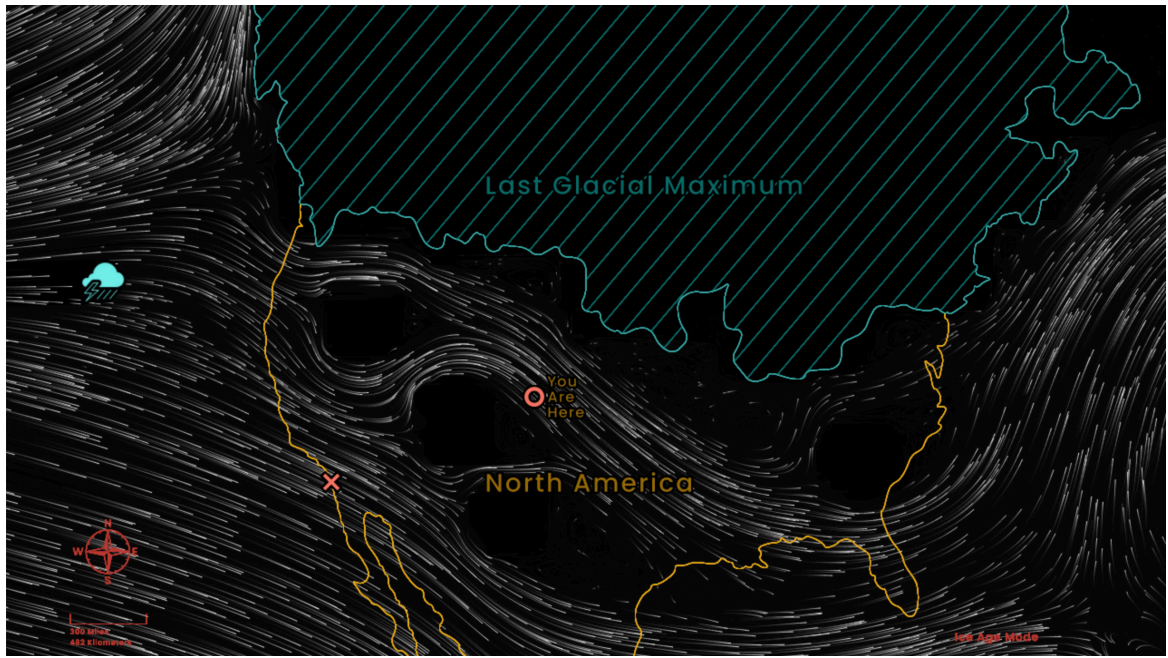

### Ice Age Mode

Visitors are challenged to recreate the LGM by placing the ice blocks and high mountain blocks in the blue shaded section the blocks. This forces the wind and storms to move south, rather than across the middle of the continent.

The map features an X target to indicate where visitors are aiming to drive storms towards, and an O target to situate the visitor at the on the map.

Wind particles are generated on the west edge and further north to visualize the influence of prevailing winds and the jet stream.

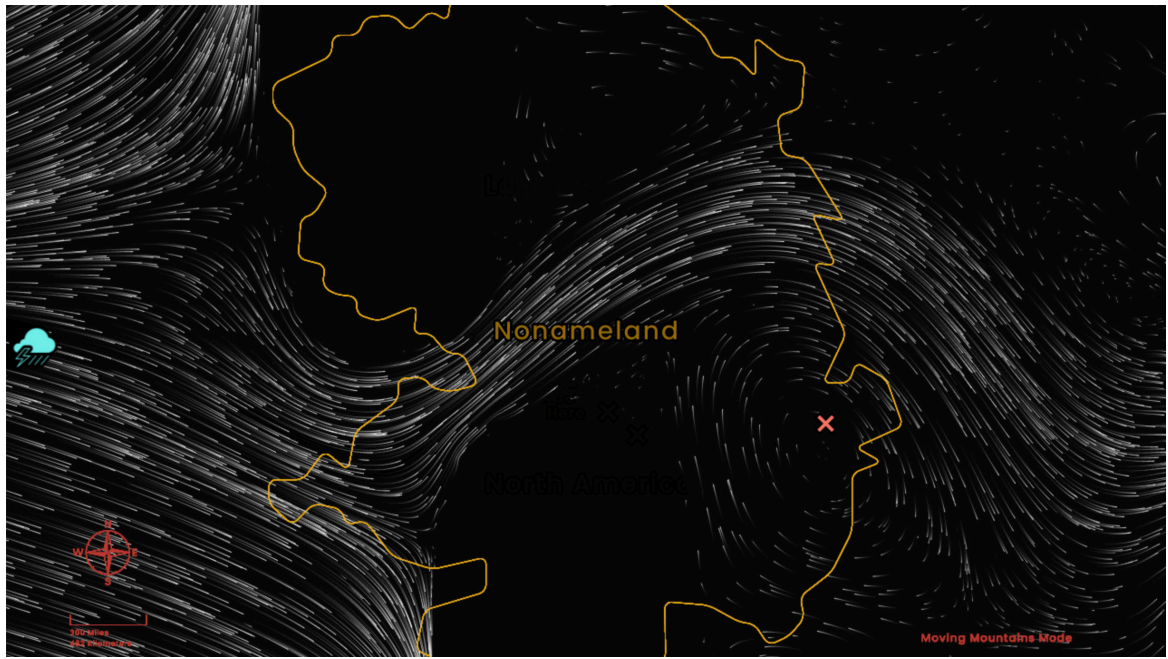

### Moving Mountains Mode

In this mode visitors are challenged to steer the wind towards an X target to learn the principles of relief effect on wind, abstracted away from a specific geographic context.

A randomly geographic shape, labelled Nonameland, is generated by selecting an outline from a range of countries and islands. The shape is then randomly rotated and flipped to further add variation for visitors.

Each time the mode is selected, Nonameland is randomly generated and the X target is placed randomly with in the central area of the map. This provides a multitude of scenarios for visitors to explore.

### Supporting Exhibit Graphics

Wall graphics were designed to provide information on the science behind the exhibit and descriptions of how radically different the climate in North America has been over time. There is a detailed explanation of how large ice sheets and mountains forced wind and wetter weather southwards in the last ice age, and the role the jetstream and ice sheets play in our climate today. It also has instructions on on interactive modes and makes thematic connections to nearby exhibits.

### Observations & Discussion: Constructing Past, Present, Future Scenarios

In the Ice Age mode visitors are explicitly recreating a past scenario, the last glacial maximum, with ice sheets and high mountains in the north to see the effects on wind and storms. Observing visitors in more detail shows that in the process of constructing this configuration (or any configuration) of land features visitors are consciously or unconsciously creating other past, present and future climate and geographic scenarios. This could be no ice sheet, like today, a smaller ice sheet in a different location, or mountains in different locations due to tectonic activity. What's more, this has a temporal component: visitors could simulate the changes in landscape across by moving the blocks co-ordinated with known geological history or then deviate to create a North American continent of the future. An open-ended simulation, even one limited to the effect of relief on wind and storms, means any time a visitor arranges the blocks they are creating a hypothetical scenario to explore what the world could look like now or "then".

### Observations & Discussion: Divergent & Convergent Thinking

The two modes encourage a problem solving mindset: how to get the wind to flow to the X target? This leads to divergent and convergent modes of thinking by first exploring what arrangement of features can move the wind closer to the target and then refining to a solution that satisfies the visitor. They are learning by doing and making. Of course they are free to ignore that goal. While the goals are explained in the supporting graphics, visitors can create their own goals (such as force the wind to their hometown) or ignore the notion of a goal and openly explore how relief changes wind and weather patterns. The simulation encourages people to ask and interact with "what if" questions they generate themselves and with others. A future version with more variables could enable users to ask more complex questions with more nuanced answers.

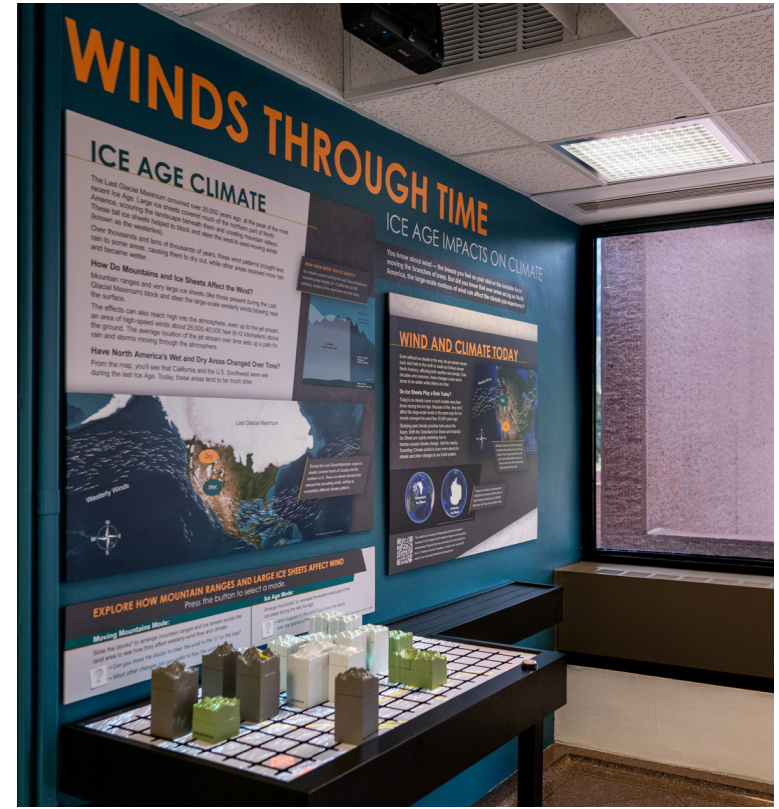

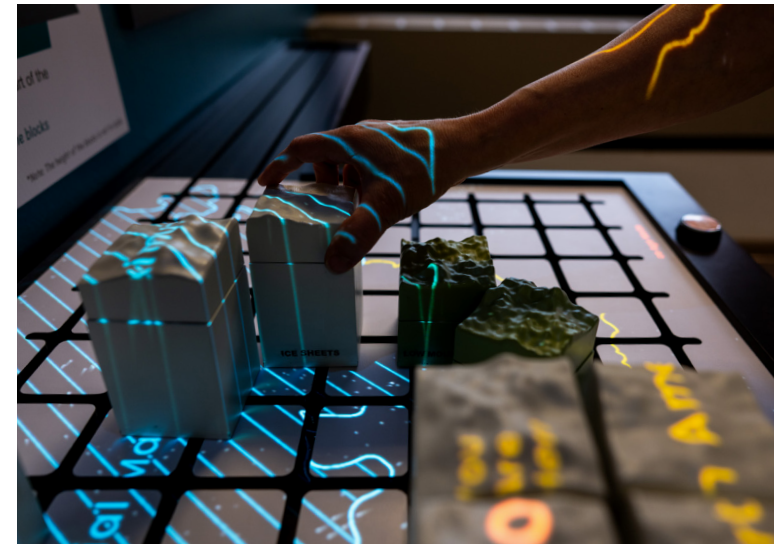

## Observations & Discussion: Shared Spaces & Resources

Observing visitors we noticed how conscious they are of the shared space and shared resources; there is a fixed tabletop and limited number of blocks. Moreover, their interactions with the simulation are dependent on each other: a visitor's interactions has an effect on the wind movements for another visitor. As a complex simulation with feedback loops everyone is downstream of each other. This contrasts with our experience of testing ARSandbox, where visitors can "claim" a section of the sandpit and choose to play independently. While the simulation is digital, the physical limitations of the exhibit and dependent interaction effects are a crucial factor in imposing this idea of a shared space where actions and resources need to be negotiated, resonating with the idea of collective care. These observations demonstrate how hybrid tangible digital data interfaces can promote user thinking towards collectivity.

## Conclusion

This project demonstrates how interdisciplinary collaboration and data visualization/physicalization can transform complex scientific research into an engaging, hands-on exhibit for the public. We created an exhibit that synthesizes data visualization and physicalization to learn about the role of relief in altering paleoclimate wind and weather through making and doing. We observed how tangible interaction and an open-ended simulation triggered a sense of shared spaces and resources, and fostered different ways of constructing and thinking about past, present, and future climate scenarios. Our work highlights the potential of design-led approaches to bridge research, education, and collective engagement.

## Acknowledgements

This work was supported by NSF (EAR 2102984)